\documentclass[journal]{IEEEtran}
\usepackage[pdftex]{graphicx}
\usepackage{graphicx}
\usepackage{url}
\usepackage{xcolor}
\ifCLASSINFOpdf
\else
\fi

\begin{document}
\title{Optimising Chest X-Rays for Image Analysis by Identifying and Removing Confounding Factors}

\author{Shahab Aslani, Watjana Lilaonitkul, Vaishnavi Gnanananthan, Divya Raj, Bojidar Rangelov, Alexandra L Young, Yipeng Hu, Paul Taylor, Daniel C Alexander, NCCID Collaborative, Joseph Jacob
\thanks{S. Aslani, B. Rangelov, Y. Hu, P.Taylor, D.C. Alexander, W. Lilaonitkul, and J. Jacob are with Centre for Medical Image Computing, Department of Computer Science, UCL Respiratory, and Institute of Health Informatics, University College London, UK.} 

\thanks{A.L Young is with Department of Neuroimaging, Institute of Psychiatry, Psychology and Neuroscience, King’s College London, UK.}

\thanks{V. Gnanananthan and R. Divya are with Department of Radiology, Royal Free London NHS Foundation Trust, London, UK.}

\thanks{NCCID Collaborative refers to all the researchers, clinicians and nurses who actively participated in contributing data to NCCID, making the creation of this database possible.}

\thanks{Corresponding author's e-mail: a.shahab@ucl.ac.uk}}

\maketitle
\begin{abstract}
During the COVID-19 pandemic, the sheer volume of imaging performed in an emergency setting for COVID-19 diagnosis has resulted in a wide variability of clinical CXR acquisitions. This variation is seen in the CXR projections used, image annotations added and in the inspiratory effort and degree of rotation of clinical images. The image analysis community has attempted to ease the burden on overstretched radiology departments during the pandemic by developing automated COVID-19 diagnostic algorithms, the input for which has been CXR imaging. Large publicly available CXR datasets have been leveraged to improve deep learning algorithms for COVID-19 diagnosis. Yet the variable quality of clinically-acquired CXRs within publicly available datasets could have a profound effect on algorithm performance. COVID-19 diagnosis may be inferred by an algorithm from non-anatomical features on an image such as image labels. These imaging shortcuts may be dataset-specific and limit the generalisability of AI systems. Understanding and correcting key potential biases in CXR images is therefore an essential first step prior to CXR image analysis. In this study, we propose a simple and effective step-wise approach to pre-processing a COVID-19 chest X-ray dataset to remove undesired biases. We perform ablation studies to show the impact of each individual step. The results suggest that using our proposed pipeline could increase accuracy of the baseline COVID-19 detection algorithm by up to 13\%. 
\end{abstract}

\IEEEpeerreviewmaketitle
\section{Introduction}
\label{Introduction}
Medical research using artificial intelligence (AI) techniques applied to clinical data and imaging is transforming our understanding of health and disease. The application of machine learning and deep learning techniques has occurred alongside an exponential growth in healthcare data acquisition. The arrival of the SARS-COV2 virus and the response of healthcare teams around the world in data collection and data sharing exemplified the scale at which medical information can be collected today for clinical research purposes. 

Large volumes of data are a prerequisite to train AI models. However, as datasets grow in size and complexity, it is important to be aware of biases that might be introduced into training AI datasets \cite{kelly2019key}. A major concern when using AI tools is that biases within training datasets may be propagated into deployed algorithms. Examples of biases include training datasets that are not representative of the target population but are imbalanced with regard to subject age, gender, ethnicity and socioeconomic or environmental factors. Such biases must be identified, understood and corrected to ensure that AI algorithms that could be used to assess patient health.

The SARS-COV2 virus has infected hundreds of millions of people across the world~\cite{who}. When the virus first emerged in late 2019, limited access to polymerase chain reaction testing kits led researchers to consider analysing medical images such as chest X-ray (CXR) and computed tomography (CT) scans to diagnose the disease and assess its severity~\cite{toraman2020convolutional,gunraj2021covid}. A hallmark of the image analysis approaches used was the reliance on deep learning systems to leverage the scale of data acquired in a timely way \cite{khan2020coronet}. Over the past two years, many automated approaches have been proposed to detect COVID-19 infection using CXR images~\cite{heidari2020improving,apostolopoulos2020covid,ozturk2020automated,khan2020coronet,rahimzadeh2020modified}. 

Though several of the proposed deep leaning-based approaches have reported excellent discriminative performance, re-evaluation of several of these methods, particularly for CXR analyses has revealed that many were confounded by non-pathological features in the images such as human annotations~\cite{degrave2021ai}. These observations highlight the importance of having a suitable processing pipeline for CXR imaging that can remove confounding signals from the image, allowing the AI model to focus on detecting and quantifying pathologically important features alone.

\begin{figure}
  \includegraphics[width=\linewidth]{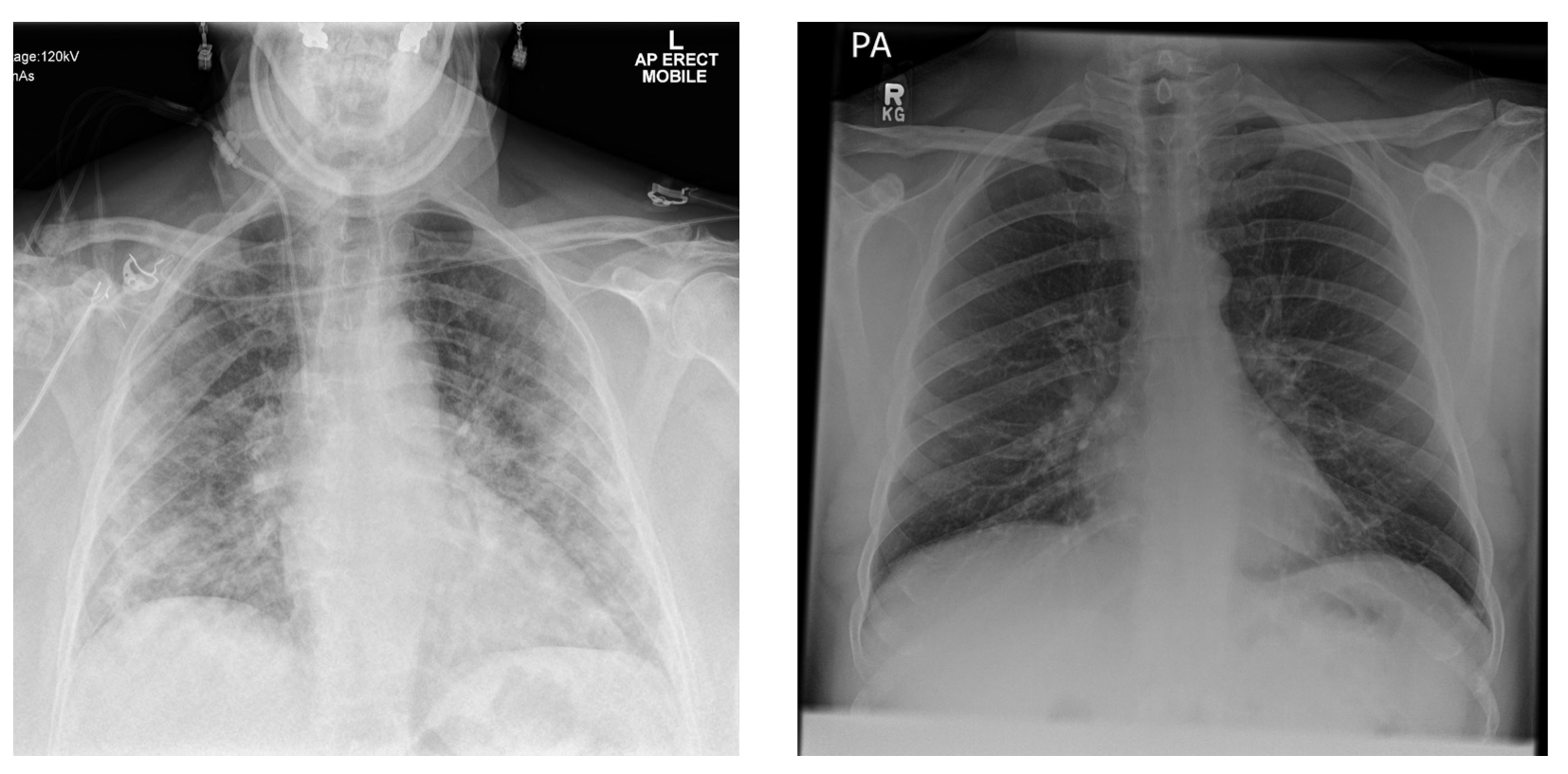}
  \caption{Anteroposterior (left) and Posteroanterior (right) chest X-ray images.}
  \label{AP/PA}
  \centering
\end{figure}

With regard to CXR projection, the optimal CXR is captured with the patient standing upright and with the x-ray beam passing from the back of the patient to the x-ray plate placed at the anterior aspect of their chest in a Posterior to Anterior (PA) acquisition. This allows the patient to perform the best expansion of their lungs in inspiration and enhances the detection of lung damage. As the heart is positioned in the anterior compartment of the chest, a PA radiograph provides a good approximation of the size of the heart relative to the size of the lungs and chest wall. A PA radiograph is the standard CXR acquisiton, however if a patient cannot stand, a radiograph can be acquired with the patient sitting or lying down and placing the X-ray plate at the patient's back, known as an Anterior-Posterior (AP) projection. In an AP projection, the size of the heart relative to the chest is exaggerated as the x-ray beams pass through the patient from front to back. AP radiographs may also be associated with a smaller inspiratory volume. Figure~\ref{AP/PA} shows an example of CXR images acquired using PA and AP projections.  All non-PA CXR acquisitions are labelled as such on the image using labels such as "AP", "sitting" or "supine". The type of CXR acquisition performed can provide information about the clinical status of a patient: a PA CXR would indicate that a patient was in better health than someone undergoing an AP CXR. COVID-19 imaging datasets typically contain a combination of AP and PA CXRs. Yet studies using AI tools to diagnose COVID-19 on CXRs invariably ignore the potential influence on algorithm performance that  may result from the algorithm simply distinguishing AP versus PA radiographs, rather than actual lung disease features.  The proportion of AP and PA CXRs also can vary across datasets, further biasing algorithms. 

Other x-ray acquisitions can also be seen in publicly available COVID-19 chest radiographic datasets. Lateral view chest X-rays (Lat-CXR) are routinely performed alongside the frontal CXR acquisition in many countries such as the USA. However in the UK lateral CXRs are rarely performed in the acute setting. If a lateral view is performed in the UK it is often to confirm or refute the presence of a lesion suspected on the frontal CXR. Therefore there is a strong bias towards the presence of pathology in cases where a lateral CXR has been performed. Abdominal X-rays (AXR) are also not uncommonly seen in COVID-19 radiographic datasets alongside the frontal CXR acquisition, particularly in patients who presented with abdominal pain. 
Having AXRs and Lat-CXRs in COVID-19 datasets can introduce some biases and image noise as these images are not useful for COVID-19 detection using AI methods. Identifying and removing these images from a dataset can help AI models improve COVID-19 detection performance.

In this paper, we describe an automated pipeline for cleaning and pre-processing CXR images. Our pipeline can help standardize a CXR dataset for future analysis. The pipeline includes: defining the unique ID per patient and per patient's CXR session; removing noisy CXR images; identifying AP and PA CXR images; and applying a lung segmentation. We show that using the pipeline to clean a COVID-19 CXR dataset can improve the performance of the baseline models for COVID-19 detection. Moreover, this pipeline will improve CXR-based deep-learning models in other existing lung diseases and for analysis of CXR imaging in potential future pandemics. The open source code will be made available soon to anyone.

\section{Materials and Methods}
\label{Materials and Methods}
\textbf{Datasets}. In this work the UK National COVID-19 Chest Imaging Database (NCCID)~\cite{cushnan2021towards,jacob2020using,cushnan2021overview} was analyzed. This dataset is a multi-center research database (comprising data from 14 NHS Hospital Trusts including 52 individual hospitals) comprising CXRs and CT scans from patients across the UK. The NCCID is a growing dataset initiated in January 2020 to enable the development of machine learning algorithms for the characterisation of patients hospitalized with COVID-19. All patients in NCCID have COVID-positive/-negative labels reflecting results of a SARS-CoV-2 RNA test via the Polymerase Chain Reaction (PCR) method. At the time of proposing our approach, 18,133 CXR images from 7,629 patients were available for analysis. 

\textbf{Method}. Since many centers were enrolled in NCCID, the imaging data, particularly the CXR imaging was highly heterogeneous with regard to imaging acquisitions. We propose an automated pipeline that can process a CXR dataset in a stepwise manner to create a standardized and homogeneous subset of the dataset, and limit potential source of bias. Our proposed pipeline includes: 
\begin{itemize}
  \item Assigning a unique ID for CXR images.
  \item Identifying and removing unnecessary/noisy X-ray images.
  \item Categorizing X-ray images into Posterior-Anterior and Anterior-Posterior acquisitions.
  \item Intensity normalization.
  \item Lung segmentation.
\end{itemize}

As a first step, we defined an image level identifier that encoded the anonymised patient identifier, the image acquisition date and the image acquisition time. This was to distinguish separate CXRs performed on an individual patient on the same day. 

\begin{figure}
  \includegraphics[width=\linewidth]{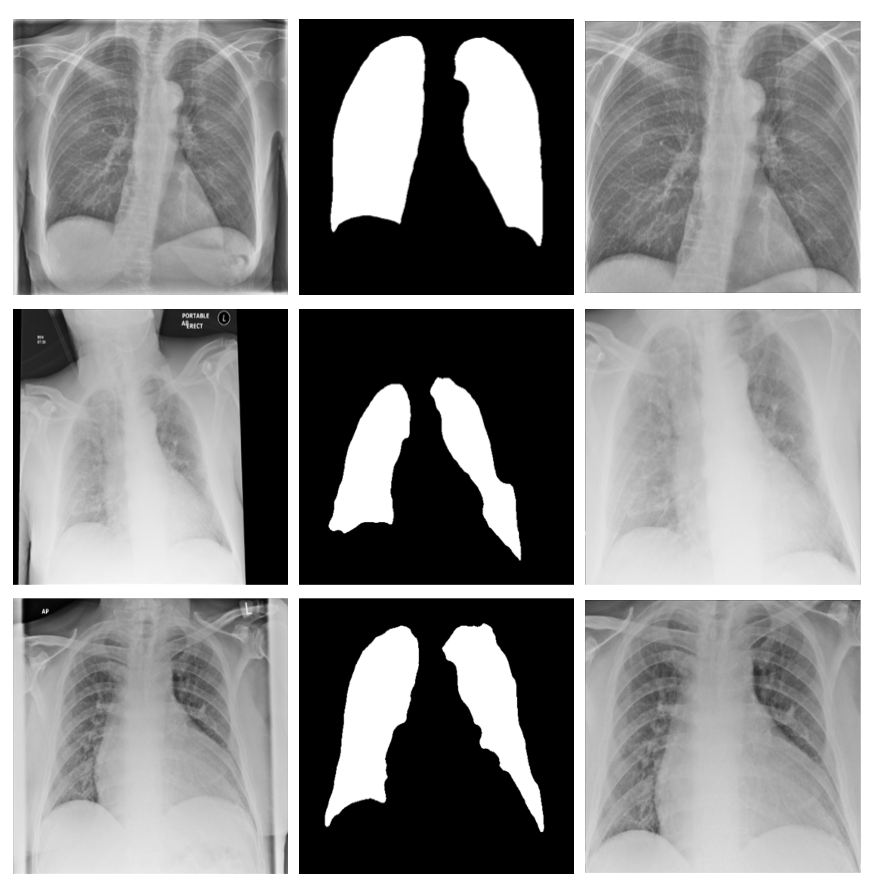}
  \caption{Lung segmentation process. We applied the lung segmentation algorithm on AP and PA X-ray images to obtain the lung masks which were used to extract the lungs from the original images (second column). We then cut the segmented images with centralizing the lung (third column).}
  \label{Seg_Out}
  \centering
\end{figure}

Within the DICOM folder of a CXR acquisition, it was not uncommon to encounter radiographs of the abdomen or lateral CXR projections. These are not suitable for COVID-19 detection. Therefore we identified them and omitted them from the dataset. Several DICOM tags were used to recognize these images automatically including ‘Series Description’, ‘View Position’, and ‘Study Description’.  

To identify AP and PA CXRs to allow for their separate processing we used the ‘Code Meaning’ sequence in the DICOM tag of the image. An example of AP and PA images can be seen in Figure~\ref{AP/PA}.

To maintain homogeneity and consistency in the dataset, all CXR images were scaled to a size of 640$\times$512. A histogram equalization algorithm was then applied to increase the contrast within the images.

\begin{table*}[t!]
\scriptsize
\centering
\caption{Output performance of our baseline EfficientNet B0~\cite{tan2019efficientnet} model for COVID-19 detection using different version of the NCCID dataset.}
\begin{tabular}{ccccccc}
\hline
Dataset & Accuracy & Sensitivity & Specificity & Precision & F1-score & AUC  \\ \hline
raw data & 0.65 & 0.56 & 0.74 & 0.68 & 0.61 & 0.72 \\
AXRs and Lat-CXRs  were removed & 0.66 & 0.60 & 0.73 & 0.69 & 0.64 & 0.75 \\
only AP CXRs & 0.73 & 0.74 & 0.73 & 0.73 & 0.73 & 0.81 \\
only PA CXRs & 0.74 & 0.74 & 0.74 & 0.74 & 0.74 & 0.81 \\
processed AP CXRs & 0.78 & 0.78 & 0.79 & 0.79 & 0.78 & 0.88 \\
processed PA CXRs & 0.77 & 0.77 & 0.77 & 0.77 & 0.77 & 0.82 \\ \hline
\end{tabular}
\label{table1}
\end{table*}

The majority of the CXR images in NCCID include text in their borders (outside of the lung area) which has been annotated manually by radiographers or radiologists. This type of information can result in undesired biases in AI models~\cite{degrave2021ai}. To remove these confounding signals, we used a lung segmentation algorithm that forced the model to concentrate on the lung region. A challenge in COVID-19 infected lungs is that they may contain peripherally placed areas of high density which can make the boundary between the lung and chest wall imperceptible. Accordingly, segmenting the lung in these cases can be a challenging task. To overcome this, we followed the idea proposed in~\cite{selvan2020lung}. The proposed segmentation architecture involves a variational encoder for data imputation, and a U-net shape network with encoder and decoder for segmentation. This model was specifically designed to segment lungs containing a high proportion of abnormalities including lung damaged by COVID-19 infection. Following successful lung extraction, we cropped the image to centralize the position of the lungs in the image.  

\begin{figure*}[t!]
  \includegraphics[width=\linewidth]{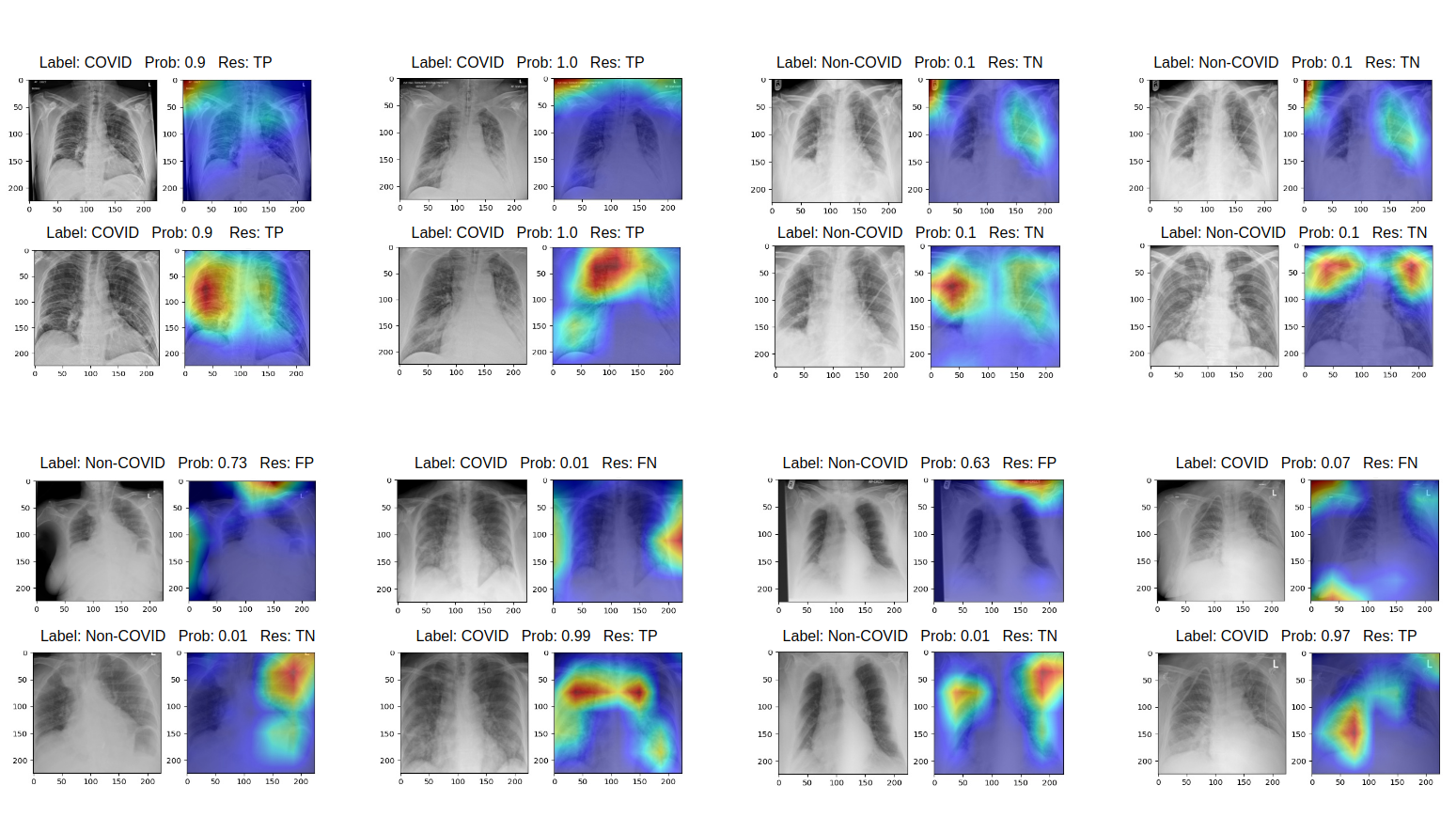}
  \caption{Baseline model output for COVID-19 detection. The first and third rows show model performance on normal CXRs including a saliency map visualization. The second and fourth rows are the model outputs on processed CXRs of the respective first and third rows. At the top of each image, information includes: 'Label'=COVID-19 ground truth label, 'Prob'=output probability of the model for COVID-1, and 'Res'=TP/TN/FP/FN outcomes.}
  \label{Saliency_maps}
  \centering
\end{figure*}

\section{Experimental Analysis and Results}
In this section, we describe experiments to highlight the robustness of the pipeline for COVID-19 detection using the NCCID dataset. We designed a deep learning-based model to generate a binary classification for COVID-19 (COVID-positive/-negative) using the PCR-based COVID-19 classification as a ground truth. A 2-dimensional EfficientNet-B0~\cite{tan2019efficientnet} pretrained on ImageNet~\cite{deng2009imagenet} was selected for our baseline model. As a first experiment, we trained and tested our model using the raw NCCID dataset to which our proposed processing pipeline was not applied. 5-fold cross validation across the entire dataset was implemented (using 18,133 CXR images) and the average of the 5-fold over test set was reported as the overall performance metric. As can be seen in the Table~\ref{table1}, the AUC performance of the model was 72\%. 

In order to gain a better understanding of the robustness of the pipeline, an ablation study was performed. We repeated the above experiment several times, iteratively processing the dataset according to the steps mentioned in our processing pipeline. In this second experiment, those images contained within CXR DICOM folders that included the abdomen and lateral CXR views were omitted from the dataset. In total 1201 images were removed from the training and testing datasets. Our baseline model AUC performance on this processed dataset was 75\% Table~\ref{table1}, a 3 percent increase compared to the unprocessed dataset. 

In the third experiment, using the aforementioned DICOM tags, we categorized the remaining images into two sub-groups (AP/PA). 71\% of the images were assigned the AP label (12023 images). 29\% were assigned the PA label (4909 images). We then repeated our cross validation experiments on AP and PA images separately. Table~\ref{table1} shows the performance of our base-line model when regarding AP and PA images separately. When analysing AP CXRs, we have a 7\% improvement in accuracy, and 6\% improvement in AUC value.

To further process the dataset, we normalized the images using histogram equalization and used the model in~\cite{selvan2020lung} to segment the lungs. Then, we cropped the images centralizing the lungs by considering the maximum and minimum values for (x,y) pixel coordinates in the extracted mask. The output can be seen in the last column of the Figure~\ref{Seg_Out}. In this experiment, we used the pre-processed AP and PA images as input to our baseline network. When using the CXRs with a segmented lung image Table~\ref{table1},  we obtained the highest AUC performance of our study: 88\% and 82\% for AP and PA images, respectively. 

We implemented a saliency map using the GradCAM~\cite{selvaraju2017grad} algorithm to confirm whether the baseline model was making COVID-19 classification decisions using pathological signals from the lung. Figure~\ref{Saliency_maps} shows some examples of the baseline model output for COVID-19 detection including the saliency map visualization using both the original CXR image and the lung segmented version of the corresponding image. The results indicate that the processed CXR helps ensure the model is making decisions based on pathology rather than confounding annotations.

All experiments were performed in the python language\footnote{\url{https://www.python.org}} using PyTorch~\cite{NEURIPS2019_9015} on a Nvidia Titan RTX 24GB GPU. We trained our model using a stochastic gradient descent optimizer with an initial learning rate of 0.001, batch size of 32, and cross-entropy as the loss function. Data augmentation was applied on the training dataset using random rotation with a maximum of 15 degrees. In all experiments, the epoch number was set to 100. We used a modified version of EfficientNet including (B0, B1, B2, and etc). We found that increasing the complexity of the model did not improve overall model performance and so EfficientNet-B0 was selected as our baseline model.

\section{Discussion and Conclusion}
We propose an automated multi-step processing pipeline to clean and standardize CXR datasets to minimise potential biases. The proposed pipeline improves dataset homogeneity and improves the performance of existing deep learning-based approaches to classify COVID-19 disease as exemplified on the UK NCCID dataset.

Defining unique identifiers for every single image in a dataset builds order into a complex dataset. Individual identifiers improve the certainty with which specific images are assigned into testing and training cohorts. When different models are then evaluated, the researcher can be confident that each model will be evaluated on identical data thereby producing a fair comparison of all available approaches. 

In a large chest imaging dataset like NCCID, which continues to expand over time, images of different body parts as well as a variety of CXR acquisitions might be captured for specific purposes. Identifying the type of images captured and categorizing them appropriately can avoid misleading AI models. For instance, in the NCCID dataset, we identified lateral projection CXRs and abdominal CXRs and removed them from the dataset as these images are unhelpful for COVID-19 detection. We showed that by simply identifying and removing these unnecessary images, the AUC performance of our COVID-19 detection model improved by 3 percent Table~\ref{table1}.

As mentioned in section~\ref{Introduction}, by separating the dataset into AP and PA CXRs, our goal was to avoid any possible undesired bias originating from the CXR acquisition in the baseline model. As shown in Table~\ref{table1}, training and testing a baseline model separately on AP and PA CXRs boosted the performance of the model with respect to all available measures including accuracy, sensitivity, specificity, precision, F1-score and AUC value.  

The first row in Figure~\ref{Saliency_maps}, shows CXRs without any processing and their model output saliency maps. All the model output predictions were correct based on the PCR ground truth and the predicted output likelihood. However, when visualizing the saliency maps, it becomes obvious that the model is basing its predictions on non-pathological signals. Specifically, the strongest predictive signals originated from text in the corner of the images, calling into question the integrity of the model predictions. In the second row, comprising segmented lungs model prediction is unchanged but now focuses on pathological signals inside the lung. 

When using deep-learning algorithms, model predictions influencing patient care may be based on imaging features unrelated to true biological damage, which could then result in patient harm. An image analysis should aim to confirm the biological plausibility of imaging features identified as having prognostic or diagnostic importance. The importance of these concepts are highlighted in the third row of Figure~\ref{Saliency_maps}, where on non-processed CXRs, the baseline model produces incorrect predictions by focusing on confounding signals outside of the lung. In the last row however, using the processed version of the images, allows the model to correctly predict outcomes using the pathological signals inside the lung.  

Whilst this study was focused on evaluation of COVID-19 CXR imaging, we believe the steps in our CXR imaging pipeline will have relevance for all researchers attempting to analyse CXR imaging no matter what the underlying disease. We also believe that such pipelines will greatly aid analysis of other respiratory infectious pathologies such as seasonal influenza outbreaks or future pandemic events.

In conclusion, this paper shows that with a simple multi-stage cleaning and processing pipeline, model performance can be boosted. We showed that categorizing CXR images systematically to avoid specific constraints associated with image acquisition can avoid biases in model prediction and result in boosted performance. We also showed that the boosted performance is based on pathological signals within the lung thereby emphasising the trustworthiness of the proposed pipeline from a clinical perspective. 

\section*{Acknowledgment}
This research was supported by Wellcome Trust Clinical Research Career Development Fellowship 209553/Z/17/Z and the NIHR UCLH Biomedical Research Centre, UK. For the purpose of open access, the author has applied a CC-BY public copyright licence to any author accepted manuscript version arising from this submission.

\ifCLASSOPTIONcaptionsoff
  \newpage
\fi
\bibliographystyle{IEEEtran}
\bibliography{ref.bib}

\begin{thebibliography}{10}
\providecommand{\url}[1]{#1}
\csname url@samestyle\endcsname
\providecommand{\newblock}{\relax}
\providecommand{\bibinfo}[2]{#2}
\providecommand{\BIBentrySTDinterwordspacing}{\spaceskip=0pt\relax}
\providecommand{\BIBentryALTinterwordstretchfactor}{4}
\providecommand{\BIBentryALTinterwordspacing}{\spaceskip=\fontdimen2\font plus
\BIBentryALTinterwordstretchfactor\fontdimen3\font minus
  \fontdimen4\font\relax}
\providecommand{\BIBforeignlanguage}[2]{{%
\expandafter\ifx\csname l@#1\endcsname\relax
\typeout{** WARNING: IEEEtran.bst: No hyphenation pattern has been}%
\typeout{** loaded for the language `#1'. Using the pattern for}%
\typeout{** the default language instead.}%
\else
\language=\csname l@#1\endcsname
\fi
#2}}
\providecommand{\BIBdecl}{\relax}
\BIBdecl

\bibitem{kelly2019key}
C.~J. Kelly, A.~Karthikesalingam, M.~Suleyman, G.~Corrado, and D.~King, ``Key
  challenges for delivering clinical impact with artificial intelligence,''
  \emph{BMC medicine}, vol.~17, no.~1, pp. 1--9, 2019.

\bibitem{who}
W.~H. Organization., ``Coronavirus disease (covid-19): Variants of
  sars-cov-2,'' p. ONLINE, 2021.

\bibitem{toraman2020convolutional}
S.~Toraman, T.~B. Alakus, and I.~Turkoglu, ``Convolutional capsnet: A novel
  artificial neural network approach to detect covid-19 disease from x-ray
  images using capsule networks,'' \emph{Chaos, Solitons \& Fractals}, vol.
  140, p. 110122, 2020.

\bibitem{gunraj2021covid}
H.~Gunraj, A.~Sabri, D.~Koff, and A.~Wong, ``Covid-net ct-2: Enhanced deep
  neural networks for detection of covid-19 from chest ct images through
  bigger, more diverse learning,'' \emph{arXiv preprint arXiv:2101.07433},
  2021.

\bibitem{khan2020coronet}
A.~I. Khan, J.~L. Shah, and M.~M. Bhat, ``Coronet: A deep neural network for
  detection and diagnosis of covid-19 from chest x-ray images,'' \emph{Computer
  Methods and Programs in Biomedicine}, vol. 196, p. 105581, 2020.

\bibitem{heidari2020improving}
M.~Heidari, S.~Mirniaharikandehei, A.~Z. Khuzani, G.~Danala, Y.~Qiu, and
  B.~Zheng, ``Improving the performance of cnn to predict the likelihood of
  covid-19 using chest x-ray images with preprocessing algorithms,''
  \emph{International journal of medical informatics}, vol. 144, p. 104284,
  2020.

\bibitem{apostolopoulos2020covid}
I.~D. Apostolopoulos and T.~A. Mpesiana, ``Covid-19: automatic detection from
  x-ray images utilizing transfer learning with convolutional neural
  networks,'' \emph{Physical and engineering sciences in medicine}, vol.~43,
  no.~2, pp. 635--640, 2020.

\bibitem{ozturk2020automated}
T.~Ozturk, M.~Talo, E.~A. Yildirim, U.~B. Baloglu, O.~Yildirim, and U.~R.
  Acharya, ``Automated detection of covid-19 cases using deep neural networks
  with x-ray images,'' \emph{Computers in biology and medicine}, vol. 121, p.
  103792, 2020.

\bibitem{rahimzadeh2020modified}
M.~Rahimzadeh and A.~Attar, ``A modified deep convolutional neural network for
  detecting covid-19 and pneumonia from chest x-ray images based on the
  concatenation of xception and resnet50v2,'' \emph{Informatics in medicine
  unlocked}, vol.~19, p. 100360, 2020.

\bibitem{degrave2021ai}
A.~J. DeGrave, J.~D. Janizek, and S.-I. Lee, ``Ai for radiographic covid-19
  detection selects shortcuts over signal,'' \emph{Nature Machine
  Intelligence}, vol.~3, no.~7, pp. 610--619, 2021.

\bibitem{cushnan2021towards}
D.~Cushnan, R.~Berka, O.~Bertolli, P.~Williams, D.~Schofield, I.~Joshi,
  A.~Favaro, M.~Halling-Brown, G.~Imreh, E.~Jefferson \emph{et~al.}, ``Towards
  nationally curated data archives for clinical radiology image analysis at
  scale: Learnings from national data collection in response to a pandemic,''
  \emph{Digital Health}, vol.~7, p. 20552076211048654, 2021.

\bibitem{jacob2020using}
J.~Jacob, D.~Alexander, J.~K. Baillie, R.~Berka, O.~Bertolli, J.~Blackwood,
  I.~Buchan, C.~Bloomfield, D.~Cushnan, A.~Docherty \emph{et~al.}, ``Using
  imaging to combat a pandemic: rationale for developing the uk national
  covid-19 chest imaging database,'' \emph{European Respiratory Journal},
  vol.~56, no.~2, 2020.

\bibitem{cushnan2021overview}
D.~Cushnan, O.~Bennett, R.~Berka, O.~Bertolli, A.~Chopra, S.~Dorgham,
  A.~Favaro, T.~Ganepola, M.~Halling-Brown, G.~Imreh \emph{et~al.}, ``An
  overview of the national covid-19 chest imaging database: data quality and
  cohort analysis,'' \emph{GigaScience}, vol.~10, no.~11, p. giab076, 2021.

\bibitem{tan2019efficientnet}
M.~Tan and Q.~Le, ``Efficientnet: Rethinking model scaling for convolutional
  neural networks,'' in \emph{International conference on machine
  learning}.\hskip 1em plus 0.5em minus 0.4em\relax PMLR, 2019, pp. 6105--6114.

\bibitem{selvan2020lung}
R.~Selvan, E.~B. Dam, N.~S. Detlefsen, S.~Rischel, K.~Sheng, M.~Nielsen, and
  A.~Pai, ``Lung segmentation from chest x-rays using variational data
  imputation,'' \emph{arXiv preprint arXiv:2005.10052}, 2020.

\bibitem{deng2009imagenet}
J.~Deng, W.~Dong, R.~Socher, L.-J. Li, K.~Li, and L.~Fei-Fei, ``Imagenet: A
  large-scale hierarchical image database,'' in \emph{2009 IEEE conference on
  computer vision and pattern recognition}.\hskip 1em plus 0.5em minus
  0.4em\relax Ieee, 2009, pp. 248--255.

\bibitem{selvaraju2017grad}
R.~R. Selvaraju, M.~Cogswell, A.~Das, R.~Vedantam, D.~Parikh, and D.~Batra,
  ``Grad-cam: Visual explanations from deep networks via gradient-based
  localization,'' in \emph{Proceedings of the IEEE international conference on
  computer vision}, 2017, pp. 618--626.

\bibitem{NEURIPS2019_9015}
\BIBentryALTinterwordspacing
A.~Paszke, S.~Gross, F.~Massa, A.~Lerer, J.~Bradbury, G.~Chanan, T.~Killeen,
  Z.~Lin, N.~Gimelshein, L.~Antiga, A.~Desmaison, A.~Kopf, E.~Yang, Z.~DeVito,
  M.~Raison, A.~Tejani, S.~Chilamkurthy, B.~Steiner, L.~Fang, J.~Bai, and
  S.~Chintala, ``Pytorch: An imperative style, high-performance deep learning
  library,'' in \emph{Advances in Neural Information Processing Systems 32},
  H.~Wallach, H.~Larochelle, A.~Beygelzimer, F.~d\textquotesingle
  Alch\'{e}-Buc, E.~Fox, and R.~Garnett, Eds.\hskip 1em plus 0.5em minus
  0.4em\relax Curran Associates, Inc., 2019, pp. 8024--8035. [Online].
  Available:
  \url{http://papers.neurips.cc/paper/9015-pytorch-an-imperative-style-high-performance-deep-learning-library.pdf}
\BIBentrySTDinterwordspacing

\end{thebibliography}

\end{document}